\documentstyle{mn}

\begin{document}

\title[H$\alpha$ searches for high redshift galaxies]
 {On near-infrared H$\alpha$ searches for high-redshift galaxies
\thanks {Based on observations collected at the European
Southern Observatory, La Silla, Chile}}

\author[A. J. Bunker {\rm et al.}]
 {A. J. Bunker,$^{1}$ S. J. Warren,$^{1}$ P. C. Hewett$^{2}$ and D. L.
Clements$^{1}$\\
$^{1}$University of Oxford, Department of Physics, Astrophysics, Keble Road,
Oxford OX1 3RH\\
$^{2}$Institute of Astronomy, Madingley Road,
Cambridge CB3 0HA}
\date{Accepted 1994 November 1.
Received 1994 September 2; in original form 1994 June 14}

\maketitle

 \begin{abstract}

The lack of success of Ly$\alpha$ searches for high-redshift $z>2$
field galaxies may be due to extinction by dust, suggesting that
surveys based on lines of longer wavelength, particularly H$\alpha$,
may be more effective. To test the dust hypothesis we have undertaken
deep broad- ($K^{\prime}$) and narrow-band (5000 km s$^{-1}$,
$\lambda=2.177~\mu {\rm m}$) imaging of the field towards the quasar
PHL957, in an attempt to detect H$\alpha$ emission from a known galaxy
of redshift $z=2.313$. We cover an area of 4.9 arcmin$^2$
(0.28$h^{-2}$~Mpc$^2$) to a 4$\sigma$ limiting narrow-band flux
$f=2.7\times10^{-16}$~erg cm$^{-2}$ s$^{-1}$, a factor of several
deeper than previously published surveys. We detect the
H$\alpha+$[N{\scriptsize~II}] emission line in this galaxy at the
$3.3\sigma$ level, inferring a star formation rate of $18 h^{-2}~
{\rm M}_\odot\,{\rm yr}^{-1}$. This is a factor only a few times larger
than the rate seen in some Sc galaxies today. The faint flux level
reached in this work demonstrates the promise of narrow-band imaging
in the near-infrared as a technique for finding normal galaxies at
high redshifts.

 \end{abstract}

\section{Introduction}

Blank-sky searches for high-redshift field galaxies ($z>2$) through the
detection of the Ly$\alpha$ emission line (de Propris et al. 1993;
Thompson et al. 1993) have so far had no success. These surveys span
the redshift range $2<z<5$, where galaxies may be experiencing their
peak star formation rate (SFR), and now cover sufficiently large areas
and reach such faint flux limits that they are in conflict with some
theories \cite{dj93}. Two competing explanations for the lack of
success are (i)  galaxies formed at higher redshift still (but note
the unsuccessful search of Parkes, Collins \& Joseph 1994), or (ii)
the Ly$\alpha$ emission is severely attenuated due to extinction by dust,
possibly exacerbated by resonant scattering from H{\scriptsize~I}
(Charlot \& Fall 1991; Valls-Gabaud 1993). Surveys based on the
detection of lines at longer wavelengths, especially
[O{\scriptsize~II}]~$\lambda3727$, H$\beta ~\lambda4861$,
[O{\scriptsize~III}]~$\lambda5007$, and H$\alpha ~\lambda6563$,
benefit from greatly reduced extinction, allowing a test of the dust
extinction hypothesis.  The H$\alpha$ line, which lies in the $K$-band
at redshifts $2.08<z<2.66$, is particularly useful because it provides
a direct estimate of the SFR \cite{ke93}.

Thompson, Djorgovski \& Beckwith \shortcite{th94} have undertaken a
pilot project for a near-infrared narrow-band imaging survey. They
used a bandwidth of 4000 km s$^{-1}$, and their deepest fields reach
$4\sigma$ detection limits in the range $1-3\times10^{-15}$~erg
cm$^{-2}$ s$^{-1}$, over a total area of 0.3 arcmin$^{2}$. Here we
report the results of a similar pilot project, using a filter of
width 5000 km s$^{-1}$, which reaches much greater depth,
$2.7\times10^{-16}$~erg cm$^{-2}$ s$^{-1}$ ($4\sigma$), and covers a
considerably larger area, 4.9 arcmin$^{2}$. We have targeted the
field of the quasar PHL957 (RA $1^{\rm h}00^{\rm m}33.4^{\rm s}$,
Dec.$=13^{\circ}00'11'', 1950.0$), $z_{\rm em}=2.681$.  Our aim was to
detect H$\alpha$ emission from the known high-redshift galaxy
in this field, $z_{\rm em}=2.313$, hereafter referred to as C1, which
was found by Lowenthal et al. \shortcite{lo91}. This galaxy lies at the
same redshift as a damped Ly$\alpha$ absorption line seen in the
spectrum of the quasar, and in fact was discovered in a narrow-band
Ly$\alpha$ search for companions of damped systems. This field
therefore is particularly interesting as it may contain other
detectable companions of these two objects.

\section{Observations}

Our broad- and narrow-band images of the field were obtained with the
IRAC2B instrument on the 2.2-m telescope at the European Southern
Observatory, over the three nights from 1993 October 31 to November 2.
Conditions were photometric, and the seeing ranged between 0.7 and 1.2
arcsec. The detector for IRAC2B is a NICMOS3 256$^{2}$ array. The
pixel scale of 0.52 arcsec pixel$^{-1}$ undersampled the seeing in
the best conditions, but gave a large field of view, corresponding to
0.5 $h^{-1}$~Mpc \footnote{We assume a cosmology with $h={\rm
H_{\circ}}/100$ km s$^{-1}$~Mpc$^{-1}$, and $q_{\circ}=0.5$ unless
otherwise stated.} comoving at the redshift of the galaxy C1.

The narrow-band filter used, ESO NB9, has a FWHM $\Delta\lambda_{\rm
na}=0.038~\mu$m, and a central wavelength $\lambda_{{\rm eff}}=2.177~
\mu$m, which closely matches the wavelength of H$\alpha$ for the
galaxy. We used the ESO $K^\prime$ filter for the broad-band
observations.  This filter is slightly narrower, $\Delta\lambda_{\rm
br}=0.32~\mu$m, than the standard $K$ filter, and shifted to a shorter
wavelength, $\lambda_{{\rm eff}}=2.15~\mu$m, thereby reducing the
sky/telescope thermal background by some 0.5 mag arcsec$^{-2}$. The
observations comprised several sequences of duration 45 min, each
made up of 9 individual 300-s frames, performed in a $3\times 3$ grid
pattern of step size 10 arcsec. Each 300-s frame was made up of a
number of co-added exposures. The integration times of the individual
exposures were 50 or 100-s for the narrow-band, and 10-s for the
broad-band, and were chosen such that the sky counts ensured the
observations were background limited, while staying well below the
saturation count level. Total integration times were 405 min for the
narrow-band, and 180 min for the broad-band. The average sky
brightness was $m_{\rm na}=12.7$, $m_{\rm br}=13.1$ mag arcsec$^{-2}$.

{}From observations of standard stars we computed extinction and colour
terms for both filters. All magnitudes quoted in this paper are in the
natural system of the filters, zero-pointed to $K$, i.e. for a star of
zero colour ($J-K=0$) $m_{\rm na}=m_{\rm br}=K$. For the colour range of the
standards used, $0.0<J-K<0.4$, the following equations apply:

\begin{equation}
m_{\rm na}=K+0.07(J-K),
\end{equation}
\begin{equation}
m_{\rm br}=K+0.13(J-K).
\end{equation}

\section{Data reduction}

A dark frame of the appropriate integration time was subtracted from
each data frame. The subsequent stages in the data reduction are
division by flat field, and sky subtraction. The number of detected
photons in an individual pixel is a sum of contributions from objects
$(O)$, the sky $(S)$, and a thermal background $(T)$ from the
telescope and instrument. Before deciding on a method of data
reduction it is important to investigate the nature of the spatial and
temporal variations of $S$ and $T$. To do this we first formed dome
flats by creating normalized frames from the difference of exposures
of the dome spot with and without illumination by a flat-field
lamp. Since $T$ is present in both frames it is removed in the
subtraction, and the flat field created in this way should be about as
good as a dome flat at optical wavelengths, i.e. an accuracy of a few
per cent on large scales, and better on small scales. For large scales
this was confirmed by an analysis of frames containing standard stars
placed at different positions within the frame. The counts of the
stars in the flattened frames show an  rms variation of
2-3 per cent.

The data frames divided by dome flats show strong gradients across
each frame, of magnitude typically 25 per cent of the mean count in the
frame, which may be imputed to the thermal background. A comparison of
frames taken at different times in the night shows that this
background varies little with time, or with mean count level. For
example the difference of two flattened frames taken 55 min apart,
during which time the mean count level varied by 15 per cent, shows an
rms  variation of 0.3 per cent of the original mean count level. (This
also provides an estimate the accuracy of the dome flat, i.e. better
than 0.3/15 or 2 per cent.) We conclude that the background in our frames
is characterized by two terms: (i) the night sky, which to first order
is spatially flat, but temporally variable, at the level of a few per
cent from frame to frame, and 20 per cent over the night, and (ii) the
thermal radiation from the telescope and instrument, which to first
order is temporally constant, but spatially variable, at the level of
25 per cent across the frame, and 1 per cent on the scale of a few
pixels. There are additional contributions to the background, at the
level of a few tenths of a per cent, in the form of arc-shaped and
v-shaped patterns whose origin is unknown. These patterns are
particularly pernicious because they exhibit both spatial and temporal
variations. Fortunately, however, they vary smoothly both in position,
on the scale of a few pixels, and in time, on the scale of a few
frames.

It is common practice to use the frames themselves to flat-field
infrared data. Since our data contain large gradients, this procedure
would introduce systematic photometric errors equal to the magnitude
of the gradients. Indeed, this was confirmed by an analysis of frames
containing standard stars, flattened using data frames. The scatter in
the standard-star photometry was considerably larger than that
achieved with dome flats. For this reason we used dome flats for
first-order flat fielding. Because the dome flats are accurate to a
few per cent, which is the desired accuracy of the final photometry, it
is irrelevant in practice whether the subsequent refinements to
produce a flat sky background in each frame involve only subtraction,
as pursued here, or a combination of division and subtraction,
i.e. first obtaining a perfect flat field, followed by sky
subtraction.

The sky, thermal, and pattern contributions were removed from each
flat-fielded data frame in three stages. First the large-scale
background was removed by subtracting a heavily median-filtered
version ($35\times35$ box) of the frame itself. The frames were then
stacked, and the median of the stack was subtracted from each
frame. This operation removes the average of the flat-field residuals
as well as small-scale variations in the thermal background, due, for
example, to dust on the dewar window. Finally a filtered
version of the data cube was subtracted, using a $5\times5\times5$
box. This process
removes the patterns, as well as any remaining background (thermal, or
flat-field residuals) that varies slowly with time. The dimensions of
the filter were the subject of experimentation to optimize the final
results, as quantified by the noise in the final combined frames, as
well as the agreement of the photometry between images obtained on
different nights. We found that a cubic box produced better results
than a one-dimensional box (e.g. $1\times1\times9$). This is because
the $x, y$ dimensions are similar to the scale of the patterns, and
(presumably) because the ratio of the dome flat and the true flat
field is smooth over these scales. Roughly speaking, the accuracy of
the flat-field and sky subtraction processing is the product of the
small-scale accuracy of the dome flat ($\sim1$ per cent per pixel) and the
short-term variability of the sky level relative to the long-term
trend ($\sim2$ per cent). This product ($\sim0.02$ per cent per pixel)
is smaller than the Poisson error for a single frame (e.g. $\sim0.1$
per cent per pixel for $K^{\prime}$).

A second approach to the flat-fielding and sky subtraction used
regression fits to the data, and solved, in an iterative manner, for
the three functions, {\it vis} the flat field $F(x,y)$, the thermal
contribution $T(x,y)$, and the sky level $S(t)$. This approach
produced very similar results.

The reduced frames were registered, scaled (to account for variations
in the airmass), and then co-added, weighting by the inverse of the
scaled sky variance, while using a percentage clipping algorithm to remove
any discrepant data points, due for example to cosmic-ray strikes. Bad
pixels flagged in a mask frame (1 per cent of all pixels) were ignored in
the co-addition.

The dithering employed means that the edge regions of the mosaicked
final images have a greater noise level, since fewer data frames
contribute than in the central region of common overlap. Therefore we
trimmed the mosaicked images to the central $256\times256$ pixel region. The
noise in the sky is uniform over most of this region, but rises
slightly towards the edges. The final broad- and narrow-band images
are shown in Fig. 1.

\section{Results}

\begin{figure*}
\vspace{22cm}
\caption[]{Narrow-band ($\lambda=2.177~\mu{\rm m}$, upper) and
broad-band ($K^{\prime}$, lower) images of the field towards the
quasar PHL957. North is up and east is to the left. The central
$256\times256$ (133  $\times$ 133 arcsec$^{2}$) region of each
mosaicked frame is shown. The broad-band frame reaches 0.8 mag deeper
than the narrow-band frame, but the object C1, a known galaxy at
$z=2.313$, is brighter in the narrow-band frame due to the
H$\alpha+$[N{\scriptsize~II}] emission line.}
\end{figure*}

To search for candidate high-redshift galaxies we performed aperture
photometry, with a radius $r=3$ pixels (1.56 arcsec), on all objects
visible in the narrow-band frame, and selected the 30 objects detected
at signal-to-noise ratio $>4$ for subsequent photometry in the
broad-band frame. The
measured magnitudes were converted to approximate total magnitudes
using an aperture correction measured from the brightest objects. It
is convenient to scale the frames to the same zero-point, such that
the relation between total apparent magnitude $m$ and counts within
the aperture $C$, for both frames, is given by

\begin{equation}
m=25-2.5\log_{\rm 10}C.
\end{equation}

The measured noise in the sky in the central regions of the scaled
frames is then $\sigma_{\rm na}=8.60$ counts pixel$^{-1}$, and
$\sigma_{\rm br}=3.95$
counts pixel$^{-1}$, which is close to the Poisson limit, and corresponds to
$m_{\rm na}=22.0$ mag arcsec$^{-2}$ and $m_{\rm br}=22.8$ mag
arcsec$^{-2}$. The 4$\sigma$ detection limit is $m_{\rm na}=19.3$, or
a narrow-band flux $f=2.7\times10^{-16}$~erg cm$^{-2}$ s$^{-1}$.

The results of the photometry are shown in Fig. 2 which plots the
$m_{\rm br}-m_{\rm na}$ colour versus narrow-band magnitude $m_{\rm
na}$. The typical colour of objects in this plot is $m_{\rm br}-m_{\rm
na}\sim 0.1$. This suggests that most of the objects are of rather
late stellar spectral type, for example cool stars, or elliptical
galaxies. Objects in the redshift range $2.29<z<2.35$ with strong
H$\alpha+$[N{\scriptsize~II}] emission lines will lie above the
sequence of normal objects. The galaxy C1 itself lies at the top of
Fig. 2. The measured magnitudes and colour for this galaxy are
provided in Table 1. The object is detected at $4.5\sigma$ in the
narrow-band, but is very faint in the broad-band ($1.9\sigma$).

\begin{table}
\caption{Properties of galaxy C1.}
\label{symbols}
\begin{tabular}{ll}

Offset rel. to PHL957 $^{\star}$ &$42.6^{\prime\prime}$W,
$24.1^{\prime\prime}$N \\
$m_{\rm na}$ & $19.22 (4.5\sigma)$ \\
$m_{\rm br}$ & $20.99 (1.9\sigma)$ \\
$m_{\rm br}-m_{\rm na}$ & 1.77 \\
Colour significance $\Sigma$ & 3.3 \\
$EW_{\rm rf}{\rm H}\alpha+$[N{\scriptsize~II}] & $>220~{\rm \AA}~(2\sigma)$ \\
$f_{\rm H\alpha}$&$2.1\pm0.6\times10^{-16}$~erg cm$^{-2}$ s$^{-1}$ \\
SFR $q_{\circ}=0.1$ & $36\pm10 h^{-2}~{\rm M}_\odot$yr$^{-1}$ \\
SFR $q_{\circ}=0.5$ & $18\pm5 h^{-2}~{\rm M}_\odot$yr$^{-1}$ \\
\end{tabular}

$^{\star}$ Offsets computed assuming the frame is oriented N -- S. The
measured position angle in our frames of star A (Fig. 1) relative to PHL957 is
$187.9^{\circ}$.
\end{table}

To select candidate high-redshift galaxies we quantify the
significance of the excess flux in the narrow-band by computing a
parameter $\Sigma$, which is the number of standard deviations between
the counts measured in the broad-band and the number expected on the
basis of the narrow-band counts. Where an object is extremely faint in
the broad-band, the error-parameter $\Sigma$ is well-defined even in
cases where the integrated counts within the registered aperture are
negative, and $m_{\rm br}$ is undefined. In computing $\Sigma$ we
assume that only the noise in the sky contributes to the errors. For
simplicity we suppose zero colour $m_{\rm br}=m_{\rm na}$, which may
be approximately correct for the continuum of a high-redshift galaxy.
Lines of constant $\Sigma$ are plotted in Fig. 2. The relation between
measured colour and $\Sigma$ is given by

\begin{equation}
m_{\rm br}-m_{\rm na}=-2.5\log_{\rm 10}[1 - \Sigma
10^{-0.4(25-m_{\rm na})}\sqrt{\pi r^{2}(\sigma_{\rm na}^{\rm 2} +
\sigma_{\rm br}^{\rm 2})} ]
\end{equation}
where $r$ is the aperture radius, 3 pixels.

\begin{figure}
 \vspace{10.5cm}
\caption[]{Colour-magnitude diagram for the 30 objects detected in the
narrow-band frame at signal-to-noise ratio $>4$. The dot-dashed
lines are lines of
constant $\Sigma$, which is the number of standard deviations of the
excess flux in the narrow-band relative to the broad-band, for an
object of zero colour $m_{\rm na}=m_{\rm br}$.  Also shown are lines
of constant rest-frame equivalent width, for the redshift of C1,
$z=2.313$.  Candidate high-redshift galaxies are objects with
equivalent widths $EW_{\rm rf}>75~{\rm \AA}$, and $\Sigma>2$. The line
$\Sigma=2$ corresponds to an SFR of $11h^{-2}~{\rm M}_\odot\,{\rm
yr}^{-1}$.}
\end{figure}

Because of the small photometric errors at bright magnitudes, objects
with only a small colour excess (either because they are red, or
because they have
lines of small equivalent width) have large values of $\Sigma$.
Therefore we need to impose an additional criterion of a minimum
equivalent width for the H$\alpha+$[N{\scriptsize~II}] line.  Lines of
constant rest-frame equivalent width $EW_{\rm rf}$ are plotted in
Fig. 2, computed from the relation

\begin{equation}
EW_{\rm rf}=\frac{\Delta\lambda_{\rm br}\Delta\lambda_{\rm na}
[1-10^{-0.4(m_{\rm br}-m_{\rm na})}]}{[\Delta\lambda_{\rm br}
10^{-0.4(m_{\rm br}-m_{\rm na})}-\Delta\lambda_{\rm na}](1+z)}
\end{equation}

For C1 we measure $\Sigma=3.3$, and $ EW_{\rm rf}=1190~{\rm \AA}$, with a
$2\sigma$ lower limit $EW_{\rm rf}>220~{\rm \AA}$, confirming that we have
detected the H$\alpha+$[N{\scriptsize~II}] line from this galaxy. The measured
${\rm H\alpha}$ line flux and estimated SFR are provided in Table 1.
The ${\rm H\alpha}$ line flux is calculated from the relation
\begin{equation}
f_{\rm H\alpha}=\frac{\Delta\lambda_{\rm br}[1-10^{-0.4(m_{\rm
br}-m_{\rm na})}]}
{(\Delta\lambda_{\rm br}-\Delta\lambda_{\rm na})}\frac
{10^{-0.4(m_{\rm na}+19.57)}}{1.33}
\end{equation}
where the first term corrects for the contribution of the continuum to
the narrow-band flux, and the second term converts from magnitude to
flux and corrects for the contribution of [N{\scriptsize~II}],
adopting the median ratio of $f_{[{\rm N~II}]}/f_{\rm H\alpha}=0.33$
found by Kennicutt and Kent \shortcite{kk83} for extragalactic
H{\scriptsize~II} regions.  To estimate the SFR we adopt the
prescription of Kennicutt \shortcite{ke83}:

\begin{equation}
{\rm SFR} = \frac{L({\rm H}\alpha )}{1.12\times 10^{41}~{\rm erg\,s}^{-1}}~
{\rm M}_\odot \,{\rm yr}^{-1}.
\end{equation}

For $z=2.313$ the following relations apply, for different values
of $q_{\circ}$:

\begin{equation}
{\rm SFR} = 1.76\times 10^{17} f_{{\rm H}\alpha} h^{-2}~{\rm
M}_\odot\,{\rm yr}^{-1}
(q_{\circ}=0.1),
\end{equation}

\begin{equation}
{\rm SFR} = 8.56\times 10^{16} f_{{\rm H}\alpha} h^{-2}~{\rm
M}_\odot\,{\rm yr}^{-1}
(q_{\circ}=0.5).
\end{equation}

In fact the SFR is proportional to the significance $\Sigma$, so the
lines of constant $\Sigma$ in Fig. 2 are lines of constant SFR. The
line $\Sigma=2$ corresponds to an SFR $=22h^{-2}~
{\rm M}_\odot\,{\rm yr}^{-1}$ for $q_{\circ}=0.1$, and to $11h^{-2}~
{\rm M}_\odot\,{\rm yr}^{-1}$ for $q_{\circ}=0.5$.

In addition to C1 there are two other objects that lie above the lines
$\Sigma=2$ and $EW_{\rm rf}=75~{\rm \AA}$ (although only just), i.e. their
colours are consistent with their being galaxies at the same redshift
as C1. The photometry for both objects, in both pass-bands, is
consistent from night to night.

\section{Discussion}

To summarize, we have undertaken a pilot study for a narrow-band
H$\alpha$ near-infrared search for high-redshift galaxies. We have
imaged a single field with an area of 4.9 arcmin$^2$, covering the
redshift range $2.29<z<2.35$.  Applying selection criteria of $
EW_{\rm rf}>75~{\rm \AA}$, SFR $>11h^{-2}~{\rm M}_\odot\,{\rm yr}^{-1}$
there are three objects in the field whose colours are
consistent with their being star-forming galaxies at the targeted
redshift. One of the objects is a previously known galaxy (C1) of
redshift $z=2.313$, for which we measure an SFR = $18 h^{-2}
{}~{\rm M}_\odot\,{\rm yr}^{-1}$. (If C1 harbours an AGN
this is an upper limit to the SFR.) This is a factor of only a few
times larger than the rate seen in some Sc galaxies today. Our
successful detection therefore demonstrates the potential of the
technique for finding normal galaxies at high redshifts. At a colder
site and with a larger telescope, such as UKIRT using the new
IRCAM3 instrument, we would reach a limiting flux a factor 2 to 3
fainter with the same integration times.

We can use our measurements of C1 to shed some light on the nature of
this object, which is a candidate primeval galaxy. Lowenthal et al.
\shortcite{lo91} measure $f_{\rm Ly\alpha}=(5.6\pm 0.1)\times 10^{-16}$~erg
cm$^{-2}$~s$^{-1}$ and $V=23.6$ for this galaxy.  For a power-law
spectral-energy distribution (SED), $f_{\nu}\propto\nu^{\alpha}$, the
measured {\it K}-band continuum flux density corresponds to
$\alpha=0.1^{+\infty}_{-0.6}$, where the limits are $1\sigma$.
Therefore the SED is consistent with the flat SED $\alpha\sim 0$
expected for a young galaxy, but a deeper $K^{\prime}$ image
is needed to place better constraints. The measured ${\rm H\alpha}$ flux
from C1 of $(2.1\pm0.6)\times 10^{-16}$~erg cm$^{-2}$ s$^{-1}$ provides
confirmation of the tentative spectroscopic detection by Hu et al.
\shortcite{hu93} who found $(2.7\pm1.2)\times 10^{-16}$~erg
cm$^{-2}$~s$^{-1}$ (where we have corrected their flux estimate for the
contribution of [N{\scriptsize~II}]). Therefore we confirm the
conclusion of Hu
et al. that extinction by dust of the Ly$\alpha$ flux from this
galaxy is fairly modest. The ratio $f_{\rm Ly\alpha}/f_{\rm
H\alpha}=2.7$ compares with the low-density Case B value of 8.3,
and implies a reddening $E(B-V)=0.16$ (computed using the
extinction law of Seaton 1979). This calculation assumes that resonant
scatterring does not significantly extend the escape path length of the
Ly$\alpha$ photons.  If, on the other hand, resonant scattering is
important, the Ly$\alpha$ line is extinguished selectively relative to
the continuum and the true rest-frame Ly$\alpha$ line EW could be
substantially larger than the measured value of $140~{\rm \AA}$. If this
were the case the object would be classified as an AGN. By measuring
the H$\beta$ line flux the true reddening could be measured, and the
intrinsic Ly$\alpha$ line EW inferred.

While the Ly$\alpha$ line in C1 is not greatly affected by dust, this
may not be true of most galaxies, given the lack of success of
Ly$\alpha$ searches for field galaxies. Therefore the faint flux level
reached in the work reported here demonstrates the promise of
narrow-band imaging in the near-infrared as a technique for finding
normal galaxies at high redshifts. One of the two other possible
high-redshift galaxies lies within the field surveyed by Lowenthal et
al. \shortcite{lo91} for Ly$\alpha$ emission, but does not show
evidence for strong Ly$\alpha$ emission. Infrared spectroscopy of
this candidate and others detected in this way will provide a
test of the hypothesis that extinction by dust is responsible for the
lack of success of surveys for high-redshift galaxies.

\subsection*{ACKNOWLEDGMENTS}

AJB and DLC acknowledge financial support from SERC/PPARC. SJW
acknowledges a Royal Society University Research Fellowship. The authors wish
to thank the staff at ESO, Chile, particularly Andrea Moneti for his
help with the IRAC2B camera.

\bsp

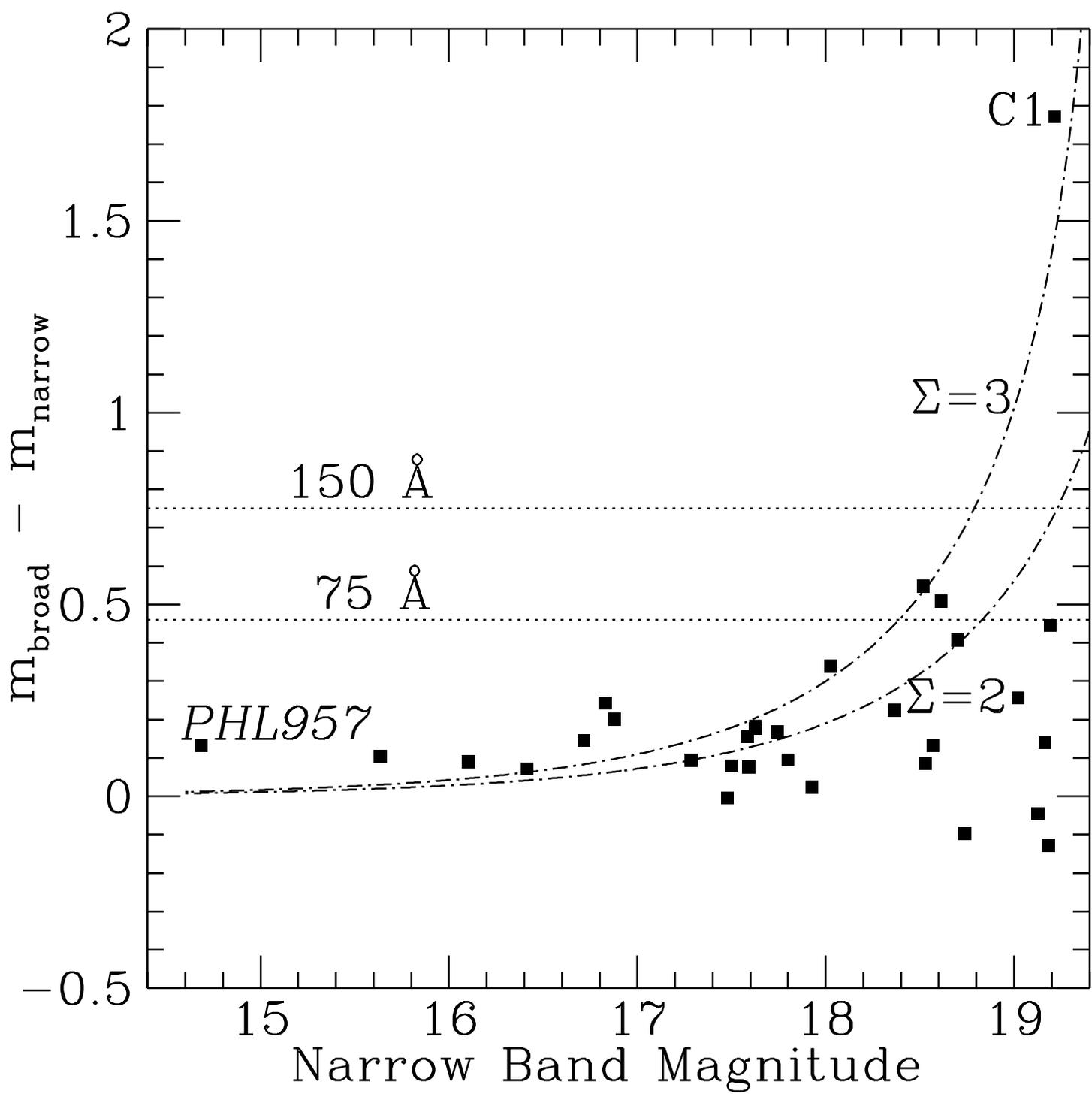
\end{document}